\documentclass{aa}
\usepackage[varg]{txfonts}
\usepackage{natbib}
\bibpunct{(}{)}{;}{a}{}{,} 
\usepackage{graphicx}
\usepackage{hyperref}
\usepackage{booktabs}
\usepackage{amsmath}

\begin{document}

\title{Convective characteristics of \ion{Fe}{I} lines across the solar disc}

\author{M. Ellwarth
  \and B. Ehmann
  \and S. Sch\"afer
  \and A. Reiners}

\institute{Georg-August Universit\"at G\"ottingen, Institut f\"ur Astrophysik und Geophysik, Friedrich-Hund-Platz 1, 37077 G\"ottingen, Germany}

\date{\today}

\abstract {Solar convection is visible as a net blueshift of absorption lines, which becomes apparent when observing quiet Sun granulation. This blueshift exhibits variations from the disc centre to the solar limb due to differing projection angles onto the solar atmosphere.}
{Our goal is to investigate convective Doppler velocities based on observations from the disc centre to the solar limb. Consequently, we aim to improve our understanding of atmospheric hydrodynamics and contribute to the improvement of solar and stellar atmospheric models.}
{We used resolved quiet-Sun spectra to investigate the convective velocity shifts of more than 1000\,\ion{Fe}{I} lines across multiple centre-to-limb positions on the solar disc. We determined the Doppler velocities with respect to the line depth. Additionally, we calculated the formation temperature and investigated its correlation with Doppler velocities.}
{The general behaviour of convective line shifts shows a decreasing blueshift as the lines become deeper for all observing positions from the centre to limb. For spectra obtained at the solar limb, even deeper lines exhibit redshifts. We observe a velocity trend for the different observation angles, with a less pronounced convective blueshift towards the solar limb. Convective velocities show a wavelength dependence for each observing angle when analysing on the basis of line depths. We observe a decreasing convective blueshift as the formation temperatures of the lines decrease. The velocity change over temperature ranges proceeds slower towards the solar limb. When investigating Doppler velocities with respect to formation temperature, the disc centre does not exhibit the strongest blueshift.
}
{}

\maketitle 

\section{Introduction}

Close-ups of the solar surface's prominent granulation pattern reveal the dynamic interplay between convective motions as remnants of the underlying convection zone. The bottom of the photosphere is covered by millions of granules all over the Sun. Intergranular lanes mark the spaces between individual granules. These granules transport plasma to the Sun's surface, where it cools and sinks back down through the intergranular lanes. The velocities of the upward-moving granules and the downward-moving plasma in the intergranular lanes are of the same order \citep[see e.g.][]{2021A&A...653A..96E}. Furthermore, it has been observed that the horizontal and vertical velocity components of the granular pattern are of equal magnitude \citep{1981A&A....96...96M}. This velocity distribution necessitates a higher density within the intergranular lanes to uphold mass conservation, and it is evident that this indeed occurs. Material density increases in radiatively cooled regions \citep[e.g.][]{1998ApJ...499..914S}.

When studying solar spectral lines from the integrated Sun or a surface area including multiple granules, a net blueshift in the Doppler velocity is observed \citep[e.g.][]{1978SoPh...58..243B, 1981A&A....96..345D}. This blueshift arises from the greater optical coverage of the surface by granules (negative, with a blueshifted velocity) compared to intergranular lanes (positive, with a redshifted velocity). Because granules appear brighter than intergranular lanes due to a temperature difference of hundreds of kelvin \citep{1998ApJ...499..914S}. This temperature contrast results in a more substantial contribution of granular photons to the overall spectrum than photons emerging from the intergranular lanes. Consequently, we observe a net blueshift when observing granulation. In higher regions of the photosphere, the granulation pattern undergoes inversion and in deep absorption lines, a reversed granulation pattern can be observed. Images captured from the line cores of the deepest spectral lines even unveil the complete absence of the granulation pattern. Depending on the specific line, deep lines may show a slight redshift in their core, even if the bisector of the line itself is blueshifted \citep{2019A&A...624A..57L}.

Observing resolved surface areas on the Sun from centre-to-limb reveals additional information about the solar atmosphere. In such observations, a prominent feature is the observed limb darkening, which accounts for a wavelength-dependent decrease in intensity from the centre towards the limb. Furthermore, the granular contrast at the limb is about three times smaller than at the disc centre \citep{1979A&A....75..223S}. \cite{1907AN....173..273H} discovered that spectral lines closer to the solar limb are redshifted compared to the lines at the disc centre. Since the observational angle between the atmosphere and line-of-sight changes from centre-to-limb, we observe changing Doppler velocities of individual lines \citep[e.g.][]{2018ApJ...866...55C, 2018A&A...611A...4L, 2019A&A...624A..57L} and different spectral properties such as limb darkening or bisectors in general \citep[e.g.][]{2019SoPh..294...63T, 2022SoPh..297....4T, 2019KPCB...35...85O, 2023A&A...671A.130P}. Light from observations closer to the limb of the solar disc pass through the atmosphere at a shallower angle, resulting in longer optical paths extending through higher atmospheric layers than observations towards the disc centre. As the optical depth increases, the line contribution function shifts to higher atmospheric layers, eventually making the line profiles  differ to the ones from lower layers \citep{2013A&A...558A..49B}.

Resolved spectral information from the Sun is important for various astrophysical fields. One field that shows keen interest in this context is the exoplanet community. \cite{2021arXiv210714291C} pointed out that stellar variability poses the most significant challenge in achieving extreme precision radial velocity capabilities for exoplanet detection. The final goal is to understand stellar signals, improve correction techniques, and to ultimately mitigate them to disentangle planetary signals from stellar signals. In their research, \cite{2019A&A...631A.100C} used 3D simulations of stellar convection and provided evidence that considering limb darkening and centre-to-limb variation noticeably improves the ability to detect exoplanets using transit spectroscopy. \cite{2017A&A...605A..90D} used 3D hydrodynamic models to simulate exoplanet transits for resolved stellar spectroscopy and pointed out the importance of testing models against resolved surface observations. Robust and detailed templates are required for this type of research. Therefore, it is important to deliver benchmark observations of resolved solar properties.

In this study, we analyse spectral observations obtained from several centre-to-limb positions on the solar disc to unravel the spatial variations in convective velocities. We employ our spatially resolved solar atlas of the Institut for Astrophysics and Geophysics (IAG) in Göttingen \citep{2023A&A...673A..19E}, which covers a broad range of the visible spectrum for multiple centre-to-limb positions. Previous studies were constrained by a limited number of spectral lines or a small number of disc positions \citep[e.g.][]{1988A&AS...72..473B, 2015A&A...573A..74S, 2019A&A...624A..57L}. The IAG atlas enables us to examine the centre-to-limb behaviour of hundreds of absorption lines. To investigate solar convection, we use \ion{Fe}{I} lines, since iron is one of the most abundant metal elements in the Sun and its absorption lines are therefore abundant and prominent in the solar spectrum. \ion{Fe}{I} lines are sensitive to temperature variations, magnetic fields, and also convective motions, making them a perfect testbed for our study. By observing the solar atmosphere from various viewing angles, we can gain insight into the depth dependence of solar convection. Understanding these convective characteristics is essential for advancing our knowledge of the underlying physical processes and their implications for stellar atmospheres.

We structure this work as followed: First, in Sect. \ref{Data} we shortly describe the observational data set and the background of the synthetically determined formation temperatures. The results of the convective variation from centre-to-limb are described in Sect.\ref{Results}, where we investigate the convective velocities depending on line depth and also on the formation temperature of the absorption lines. Eventually, in Sect.\ref{Summary}, we summarise our results.

\section{Data} \label{Data}

We used spatially resolved solar spectra to investigate \ion{Fe}{I} absorption lines from the centre to the limb. Additionally, we model the temperature formation height of those lines to examine the respective convective velocities.

\subsection{Observed spectra}
 
We analysed spectra from our IAG spatially resolved quiet Sun atlas, examining nine distinct centre-to-limb positions. We utilised observations ranging from $\mu$ = 1.0 at the disc centre towards the solar limb at $\mu$ = 0.2 in steps of 0.1. The positions of the observation relative to the centre are defined as $\mu = \cos \theta$, where $\theta$ is the heliocentric angle of the Sun. These observations were obtained using the 50\,cm Vacuum Vertical Telescope located at the IAG in Göttingen. The spectral data was acquired using our Fourier Transform Spectrograph (FTS), \citep{2020SPIE11447E..3QS}, where we chose a spectral resolution of $\Delta \nu$ = 0.024~cm$^{-1}$, corresponding to a resolving power of $R = \nu/\Delta \nu \approx$ of 700,000 at $\sim$6000\,\r{A}. A comprehensive description of the telescope and the instrumental set-up can be found in \cite{10.1117/12.2560156}. We estimated that the uncertainty in the absolute zero point offset of each spectrum is to be less than 30\,m\,s$^{-1}$ \citep{2023A&A...673A..19E}. We applied a net velocity correction to correct for the gravitational redshift of $\Delta v$\,=\,$-$633\,m\,s$^{-1}$. Our study focuses on the spectral behaviour of over a 1000 of \ion{Fe}{I} lines within the spectral range between $4300-7500$\,\r{A}.

\subsection{Formation temperature} \label{pysme}

To deepen our understanding of the physical conditions and processes contributing to spectral lines, it is beneficial to know the temperature and pressure conditions in the layers of the solar atmosphere where these lines are formed. Line depths of absorption lines do not provide accurate information about the formation height in the solar photosphere. Employing the lines' formation temperature as a proxy for formation height allows us to arrange the lines by altitude in the solar atmosphere.
In this study, we estimated the formation temperature of the absorption lines using the spectral synthesis code Spectroscopy Made Easy (SME), \citep{1996A&AS..118..595V, 2017A&A...597A..16P}, which provides a reliable method for determining an average of line formation temperature. We used PySME\footnote{\url{https://github.com/AWehrhahn/SME}}, a python implementation \citep{2023A&A...671A.171W} of SME, to perform the synthesis. In order to model a 1D local thermodynamic equilibrium (LTE) atmosphere using PySME, we utilised the atomic line list provided by the Vienna Atomic Line Database (VALD3, \cite{1995A&AS..112..525P,2000BaltA...9..590K,2015PhyS...90e4005R}) and MARCS model atmospheres \citep{2008A&A...486..951G}. We chose the solar parameters for our simulations, with an effective temperature of $T_{\text{eff}} = 5778 K$ and a surface gravity of log $g = 4.437$.

PySME offers users a temperature profile across a grid of optical depths ($\tau$) for the reference wavelength of 5000\,\r{A}. Additionally, it provides parameters such as density, $\rho_\nu$, continuum opacity, and line opacity, with the latter two summing up to the total opacity $\kappa_\nu$. These parameters enable the calculation of optical depths for other wavelengths using:
\begin{align}
    \tau_\nu = \int_{0}^{D} \kappa_\nu \rho_\nu \text{d}s,
\end{align}
where $D$ represents the thickness of the PySME grid, and d$s$ is the path through the grid. To calculate the contribution function at specific positions ($\mu$) over the solar disc, we employed a plane-parallel surface approximation. This approximation is suitable because the photosphere constitutes only $\approx 0.1\,\%$ of the solar radius, making the curvature's radius much greater than its thickness. The contribution function, containing the emergent flux as a function of $\mu$, is calculated by:
\begin{align}
    CF(\tau_\nu, \mu) = \int_{0}^{\tau_\lambda} S_\nu e^{-\tau_\nu/\mu} \frac{1}{\mu}d\tau_\nu. \label{eq:contrib}
\end{align}
The temperature profile and the contribution function share the same optical depth grid for any given wavelength. Consequently, we can derive the contribution function as a function of temperature. Since spectral lines form across a vast temperature range, we decided to utilise the average line formation temperature on the PySME grid, represented by the temperature at which 50\,\% of the flux for the specific wavelength is emitted.

\section{Results} \label{Results}

In this section, we present our findings regarding the overall behaviour of the core positions of iron \ion{Fe}{I} absorption lines across the solar disc. We particularly focus on the change of convective velocity over line depth and formation temperature from the centre to limb.

\subsection{Convective blueshift over line depth} \label{lindepth}

\begin{figure*}
        \centering
        \includegraphics[width=0.9\textwidth]{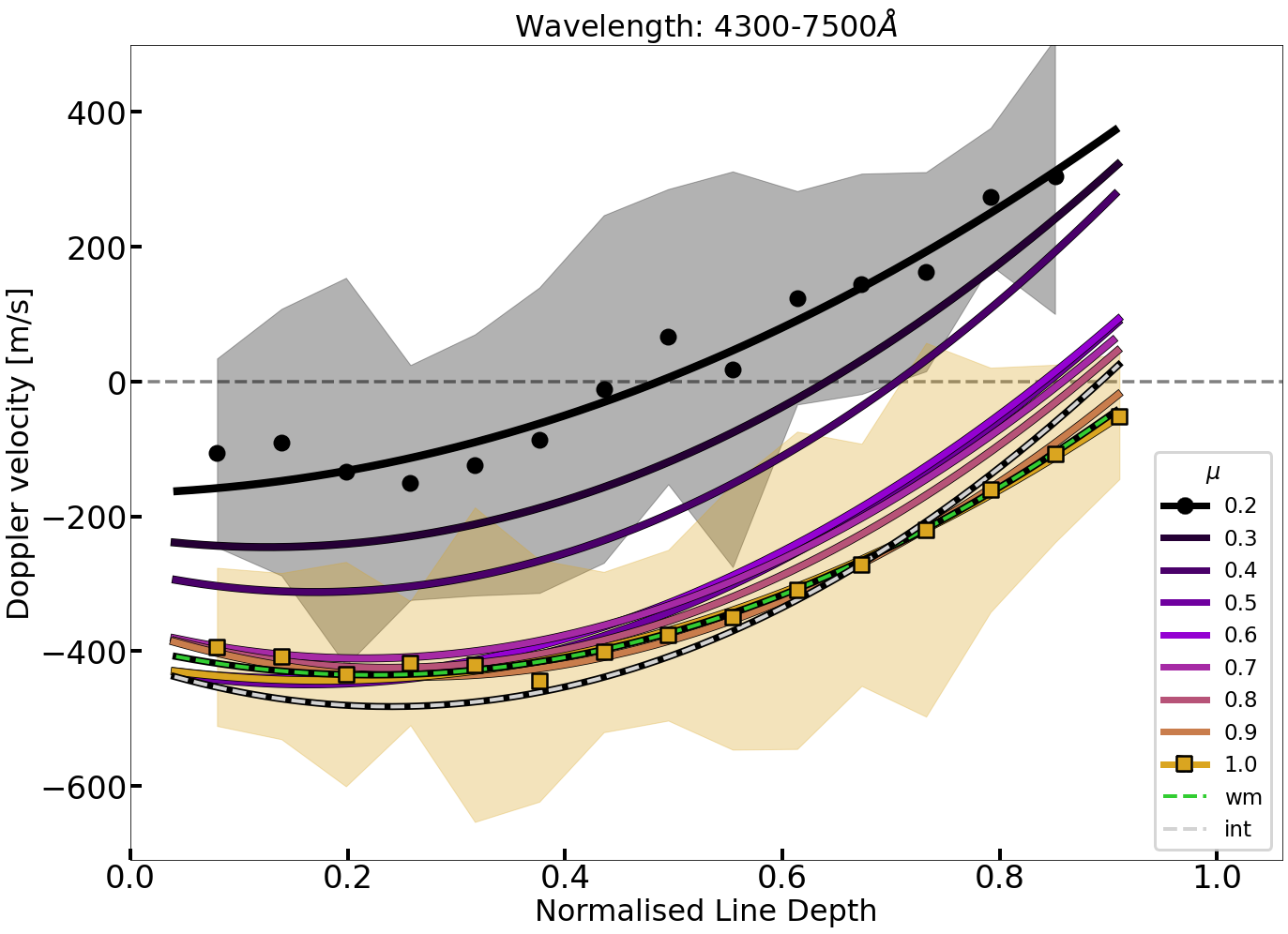}
        \caption{Fitted Doppler velocity as a function of line depth of \ion{Fe}{I} lines for various $\mu$ positions. The weighted mean ('wm') of these $\mu$ positions is also displayed. Velocities obtained from the integrated IAG atlas ('int') are presented for comparison. For $\mu$\,=\,1.0 and $\mu$\,=\,0.2, we show binned data points of line depth of chunks of 0.06. The shaded areas represent the standard deviation of the bins for those $\mu$s.}
        \label{fig:dep_veloall}
\end{figure*}

As a reference for our analysis, we utilised rest wavelengths from the VALD3 database \citep{K14} and used the unblended iron line list of \cite{1989A&AS...77..137T, 1990A&AS...82..179T} to exclude blended lines from the VALD3 list. Within our wavelength range of $4300-7500$\,\r{A}, we identified a total of 1055 \ion{Fe}{I} lines for the observation at the disc centre. The number of identified lines slightly decreases towards the limb, with 1012 iron lines identified at $\mu$ = 0.2. To determine the corresponding Doppler shift between rest wavelengths and the spectral line positions from the IAG atlas, we identified the centre of each line by fitting a parabola to the line core within a range of 0.08\,\r{A}. Figure~\ref{fig:dep_veloall} shows the resulting Doppler velocities at various $\mu$ positions over the normalised line depth. For clarity, we do not display the velocity results of every individual absorption line; instead, we apply second-order polynomial fits to reveal trends of every $\mu$. The corresponding coefficients for these fits, as well as the average standard deviation between fits and data points, can be found in Table~\ref{tab:poly}. For $\mu$\,=\,1.0 and 0.2, we additionally bin the velocity into line depth chunks of 0.06 to provide a better sense of the data and visualise the standard deviation. We only show bins for these two $\mu$s, to keep the plot neat. The standard deviations are of the order of about 220\,m\,s$^{-1}$. We notice a broad velocity scatter across the lines in general, however, about 95\% of the data points are distributed within the 2\,$\sigma$ range around the fit. The data exhibit a generally normal distribution including a few outliers. A significant source of uncertainty originates from the determination of the line core position itself. When fitting a parabola to the line core, we use this polynomial as an approximation. However, this approach will never fully capture the true nature of an absorption line, as our assumptions are inherently simplifications. Moreover, line cores may exhibit tilts, introducing additional uncertainty when determining the true line core. Since we use about 1000 \ion{Fe}{I} lines per $\mu$, we assume this measurement uncertainty to be balanced out. Another reason for the wide scatter is the wavelength dependence of the Doppler velocities, which we will explore in more detail further on in this paper.

\begin{table}
        \centering
        \small
        \caption{Coefficients of a second-order polynomial.}
        \begin{tabular}{@{}rrrrr@{}}
                \hline
                \hline
                $\mu$ & a & b & c & $\sigma$ [m/s]\\
                \hline
                0.20 & 603.5086 & 47.4886 & -165.6558 & 225 \\
                0.30 & 927.5267 & -234.2827 & -230.7359 & 226\\
                0.40 & 1089.4243 & -370.9079 & -280.6027 & 247\\
                0.50 & 967.0701 & -316.2285 & -423.2393 & 249\\
                0.60 & 917.2825 & -269.9562 & -420.1709 & 245\\
                0.70 & 989.5218 & -423.1094 & -365.3772 & 240\\
                0.80 & 1044.1613 & -495.5345 & -365.8733 & 231\\
                0.90 & 1012.5575 & -539.0614 & -366.0221 & 207\\
                1.00 & 728.4024 & -257.2781 & -420.6784 & 192\\
                \\
                wm & 837.0296 & -373.4604 & -393.5803 &  \\
                int & 1127.7019 & -539.4649 & -417.5543 & 256 \\
                \hline
        \end{tabular}
    \tablefoot{ Coefficients of a second-order polynomial $\Delta v = ax^2 + bx + c$ as plotted in Fig.\,\ref{fig:dep_veloall} and the averaged standard deviation between the fit and the data points.}
\label{tab:poly}
\end{table}   

The Doppler velocities of iron lines at the solar disc centre differ from those at the furthest solar limb position ($\mu$=0.2) by an average offset of up to 350\,m\,s$^{-1}$. The absorption lines at limb positions are more redshifted than at the disc centre. Observing at the solar limb includes observing higher atmospheric layers. This trend, which indicates an increasing redshift of the lines for higher atmospheric regions, could already hint at the redshift of chromospheric UV lines \citep[e.g.][]{1999ApJ...522.1148P}. Generally, the observed convective blueshift decreases as we move away from the disc centre with a notably strong decrease after $\mu$\,<\,0.5 \citep[cf.][]{2023A&A...673A..19E, 2019A&A...624A..57L}. However, the fit for the Doppler velocities at $\mu$\,=\,0.5 and $\mu$\,=\,0.9 shows a slightly stronger convective blueshift than the disc centre at line depths between 0.05 to 0.3 and 0.3 to 0.7, respectively. The most pronounced blueshifts are observed for the shallower lines at positions of $\mu$\,>\,0.5. As we move towards deeper lines, the velocity increases for every $\mu$.
For the disc centre the mean velocity of the lines is at its lowest at a depth of 0.2 with $-440$\,m\,s$^{-1}$, increasing to $-2$\,m\,s$^{-1}$ at a line depth of 0.95. It is expected that weak lines will exhibit a stronger blueshift \citep{1981A&A....96..345D,1986A&A...158...83D}, as they emerge at the lower end of the photosphere where granulation is most pronounced. For the limb position of $\mu$\,=\,0.2, the velocity for shallow lines is approximately $-150$\,m\,s$^{-1}$, increasing with line depth to about 350\,m\,s$^{-1}$. At $\mu$\,=\,0.2, the resulting Doppler velocity transitions from a blueshift into a redshift beyond a line depth of approximately 0.5. Similarly, for $\mu$\,=\,0.3 to $\mu$\,=\,0.8, the deepest lines exhibit redshifted velocities, while closer to the disc centre their average velocities remain blueshifted.

All $\mu$ positions show the same trend of increasing velocities with increasing line depth. The transition from blueshift to redshift occurs at progressively deeper spectral lines for observations closer to the disc centre. This trend occurs because observations at the disc centre do not probe the same geometric height in the solar atmosphere as observations towards the limb. As the observational angle continuously rises, we observe higher altitudes above the solar surface. The convective blueshift is a result of surface granulation, which diminishes as the atmospheric velocities become less correlated to this pattern further away from the granules. Hence, it is unsurprising that the convective shift decreases for higher atmospheric layers. The effect of a redshifted solar limb was previously identified by \cite{1985SoPh...99...31B}. The author explained this redshift by the different geometric depths of log $\tau$ = 0 for granules and intergranular lanes.


In order to investigate the differences between the results obtained from resolved Sun observations with those from the integrated Sun, we additionally consider the data of the integrated IAG solar flux atlas from \cite{2016A&A...587A..65R}. This comparison also serves as a valuable cross-validation of our findings, as the velocities derived from the integrated atlas data are anticipated to correspond closely to the disc centre. \cite{2016A&A...587A..65R} calculated the convective blueshift velocities of the integrated solar \ion{Fe}{I} lines and provided a polynomial fit to these results. For consistency, we first re-calculated the Doppler velocities from the integrated Sun observations using the VALD3 rest wavelengths. We find an offset of about 45\,m\,s$^{-1}$ towards higher velocities (less convective blueshift) between the fit of our re-calculated velocities and the original fit. We ascribe this offset mainly to different rest wavelengths. \cite{2016A&A...587A..65R} calculated the Doppler velocities using the line list of \cite{1994ApJS...94..221N}. For validation, we compared the line lists of VALD3 and \cite{1994ApJS...94..221N}. Therefore, we calculated the corresponding velocities between these line lists and found an offset of approximately 31\,m\,s$^{-1}$. Further details about the offset calculation can be found in Appendix \ref{deviation}. We include the convective blueshift velocities for the integrated atlas into Fig.~\ref{fig:dep_veloall}. The fits of the integrated data and the disc centre ($\mu$\,=\,1.0) fall within the range of each other's standard deviation.

For further comparison, we computed the weighted average of the Doppler velocities across the nine $\mu$ positions, which is also depicted in Fig.~\ref{fig:dep_veloall}. For this weighted average, we take the disc position and limb darkening over the solar disc \citep{1994SoPh..153...91N} into account. This weighted mean is expected to portray the integrated data. As seen in the plot the polynomials differ slightly in terms of strength and slope. The weighted average is close to the velocities of the disc centre and nearly overlapping with those for the line depth area between 0.3 and 0.8. In comparison with observations of the Sun as a star, the weighted average exhibits up to 70\,m\,s$^{-1}$ more blueshifted for stronger lines, and is about 40\,m\,s$^{-1}$ less blueshifted for shallower lines than 0.5.
\begin{figure}[ht]
    \centering
    \includegraphics[width=0.49\textwidth]{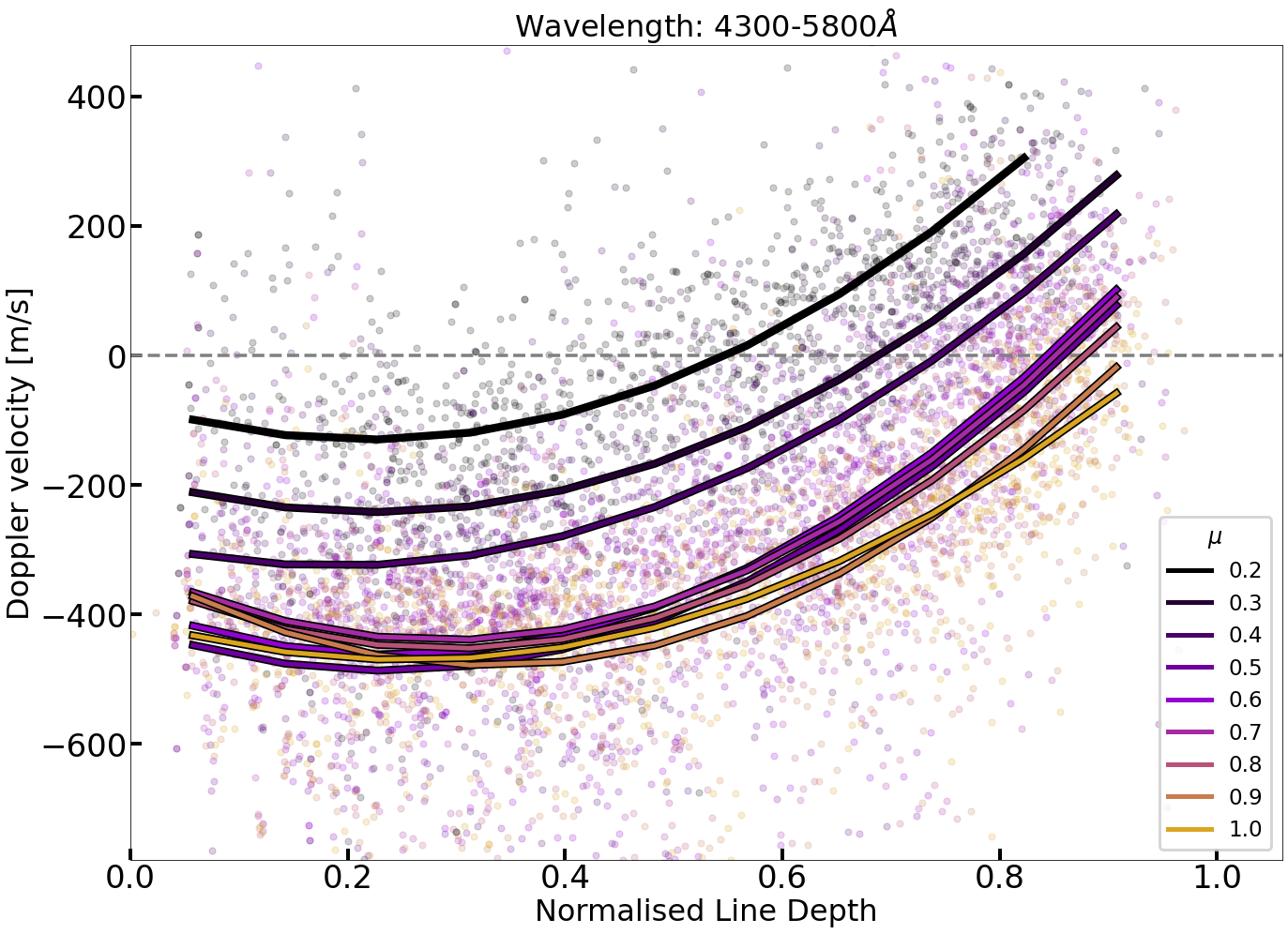}
    \includegraphics[width=0.49\textwidth]{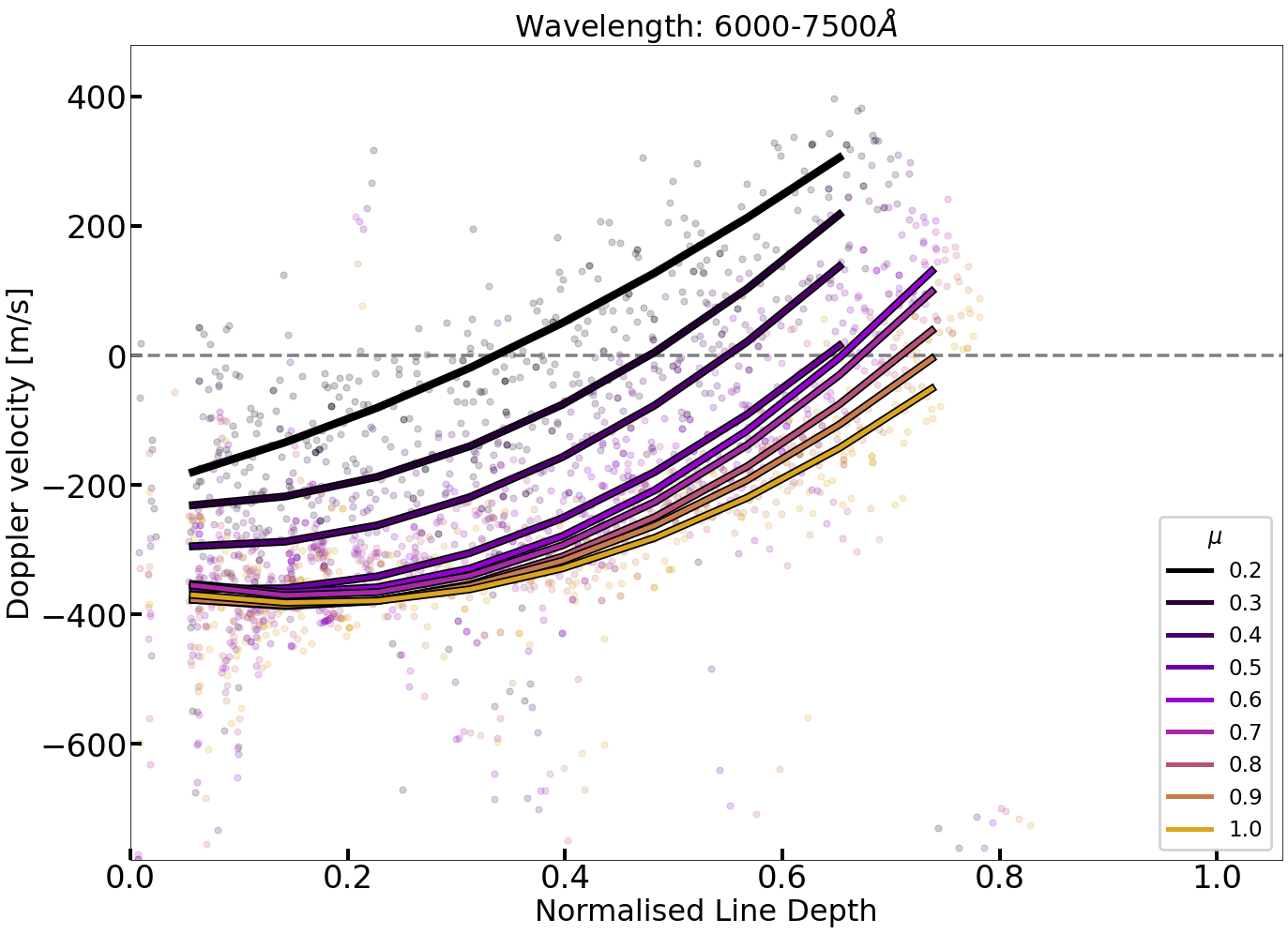}
    \caption{Doppler velocity as a function of line depth  of \ion{Fe}{I} lines, for various $\mu$ positions. The dots show the individual Doppler velocities of the lines while the lines show the corresponding fits to each $\mu$. Upper: Wavelength range $(4300-5800)$\,\r{A}. Lower: Wavelength range $(6000-7500)$\,\r{A}.}
    \label{fig:dep_velo_b}
\end{figure}  

The strength of absorption lines varies across different sections of the spectrum, showing changing maximum depths. The most pronounced \ion{Fe}{I} lines are found in the spectrum's blue part, while the line depth in the red part shows weaker lines in general. Absorption lines at shorter wavelengths show a larger convective blueshift because the temperature difference between the hot rising granules and the cooler downflowing intergranular lanes cause a greater stellar surface contrast at this wavelength range due to the higher energy \citep[see][]{2000A&A...359..729A}. This wavelength-dependency of line depths prompts us to investigate variations in Doppler velocities across different wavelength ranges. In Fig.~\ref{fig:dep_velo_b}, we show the Doppler velocities and its polynomial fits as a function of line depth for two wavelength ranges: $4300-5800$\,\r{A} and $6000-7500$\,\r{A}. It is evident that, in general, the lines exhibit greater depths at shorter wavelengths. Between $4300-5800$\,\r{A}, lines can reach depths of more than 0.9, while in the longer wavelength range of $6000-7500$\,\r{A}, the maximum line depths are around 0.75. When examining identical line depths, we observe distinct velocity variations between the two wavelength ranges. For instance, at a line depth of 0.5, the fits at the shorter wavelength range exhibits only blueshifted velocities for all $\mu$  values (below zero), whereas the polynomials of the longer wavelength range already show a redshifted effect in the limb observations. 

\begin{figure}
        \centering
        \includegraphics[width=0.49\textwidth]{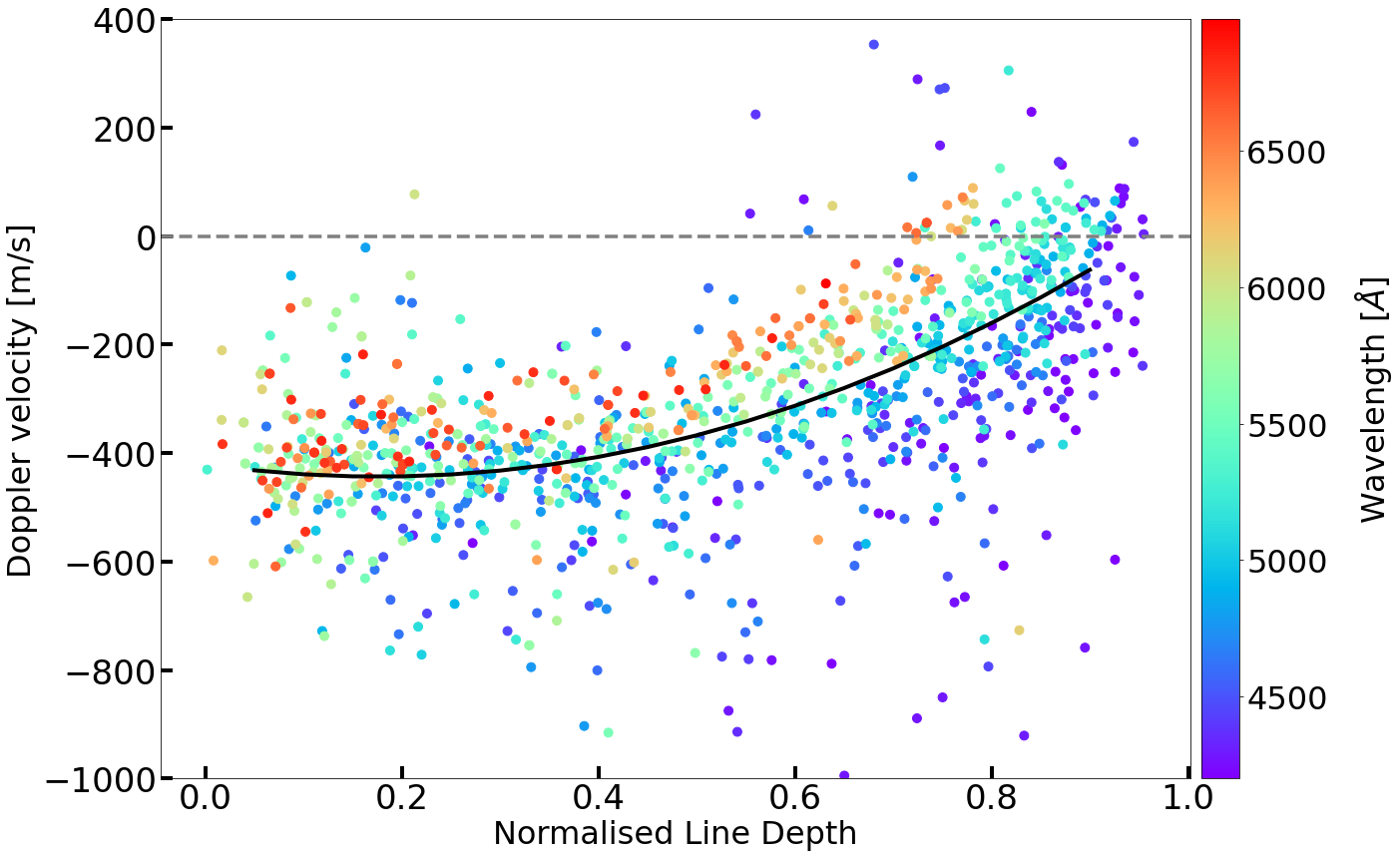}
        \caption{Doppler velocity of \ion{Fe}{I} line cores for $\mu$ = 1.0, as a function of line depth. The results are colour coded over wavelengths. The black line corresponds to the fit from Table~\ref{tab:poly}.}
        \label{fig:velo_depth_wl}
\end{figure}

To evaluate the wavelength dependency in more detail, we employed the spectrum at the disc centre $\mu$ = 1.0 for a comprehensive analysis. Figure~\ref{fig:velo_depth_wl} 
presents an example of the convective velocities' distribution over varying line depths, colour-coded for the changing wavelengths. The velocity trend changes depending on wavelengths and lines at shorter wavelengths show generally deeper lines. This wavelength-dependent maximal line depth change occurs from mainly two origins. On one hand, photons with shorter wavelengths are more effectively absorbed in the atmosphere, resulting in stronger absorption lines in this wavelength range. On the other hand, it is important to consider the influence of various opacity sources in the solar atmosphere. One major role is played the H$^-$ continuum's opacity. This absorption coefficient increases towards larger wavelengths of our wavelength range \citep{1946ApJ...104..430C}. A larger absorption coefficient means a shortened path length of the emitted light and therefore the depth of the photosphere decreases \citep{2005oasp.book.....G}. Consequently, a higher continuum's opacity leads to shallower lines.

\begin{figure}
        \centering
        \includegraphics[width=0.49\textwidth]{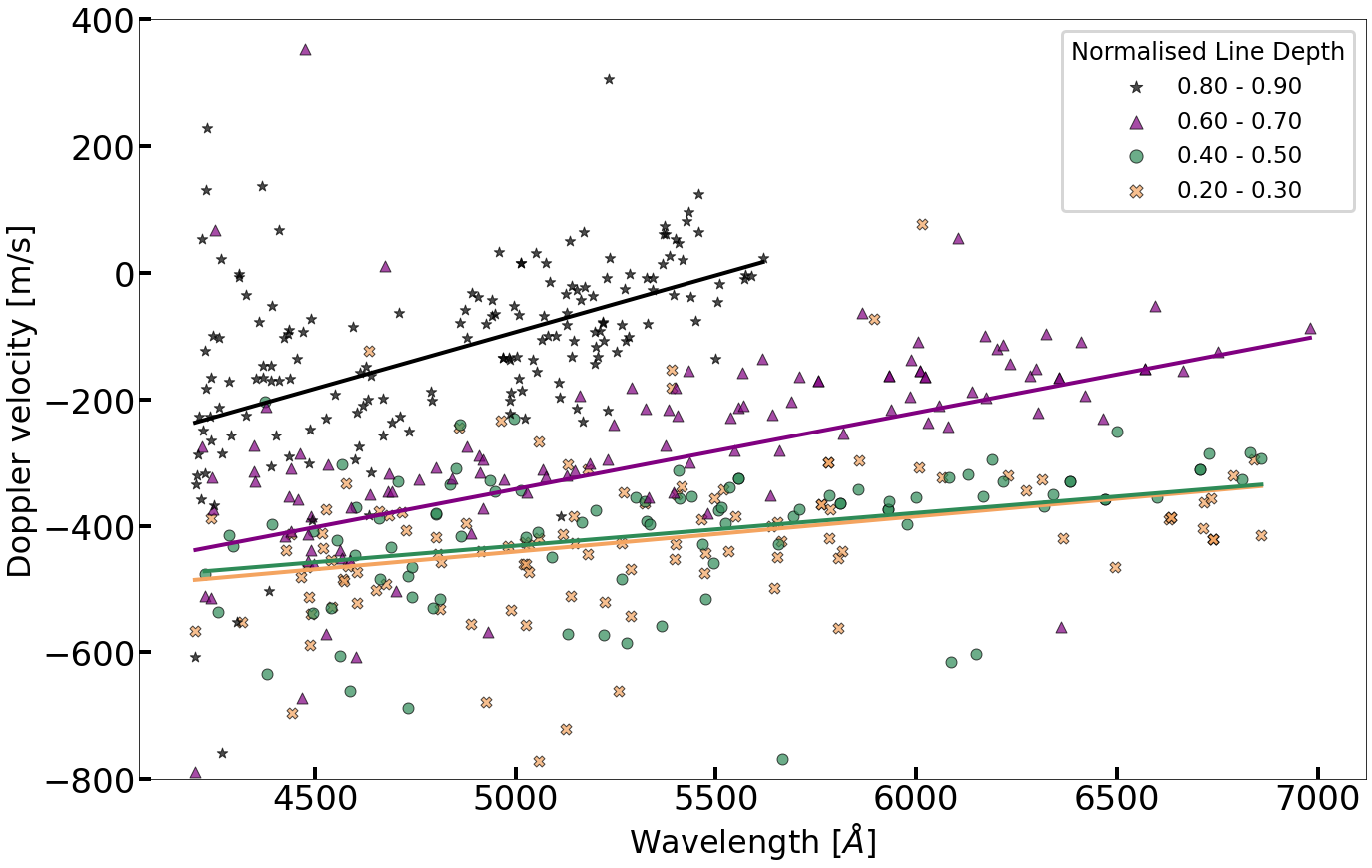}
        \caption{Doppler velocity of \ion{Fe}{I} line cores for $\mu$ = 1.0 plotted as a function of wavelengths, for four line depth intensity chunks. The higher the number the deeper the line.}
        \label{fig:wavel_velo}
\end{figure}

We go on to analyse the wavelength profile of the data from Fig.~\ref{fig:velo_depth_wl}. We divided the line depth in chunks and plotted the convective blueshift as a function of wavelengths which is illustrated in Fig.~\ref{fig:wavel_velo}. The behaviour over wavelength differs significantly among the different chunks, which can be highlighted by fitting linear functions. Without considering depth, the velocity distribution appears uncorrelated over wavelength, as shown by \cite{1999PASP..111.1132H}. The authors used a solar flux atlas and showed also the dependence on line depth, finding a similar outcome as in this work. For the deepest lines we find a convective blueshift between -220\,m\,s$^{-1}$ (at 4200\,\r{A}) and nearly zero (at 5500\,\r{A}). The trend of stronger convective blueshift towards shorter wavelengths is noticeable in each depth chunk, but with slightly different slopes. We notice a tendency between line strength and wavelength dependency of those lines. The stronger the lines, the stronger also the wavelength dependence for the line depth chunks.
Examining the $\mu$ positions towards the limb, the velocity dependence on wavelength persists, albeit with varying offsets and slopes. The trend of stronger dependence for stronger lines is also noticeable here. This evaluation proofs wavelength dependence regarding to line depth and that the influence even changes depending on the depth of the lines itself. To get a more detailed picture of the solar atmosphere and the convective velocities, it is crucial to consider this wavelength dependencies.

\subsection{Convective blueshift over mean formation temperature}
\label{lintemp}

\begin{figure*}
        \centering
        \includegraphics[width=0.9\textwidth]{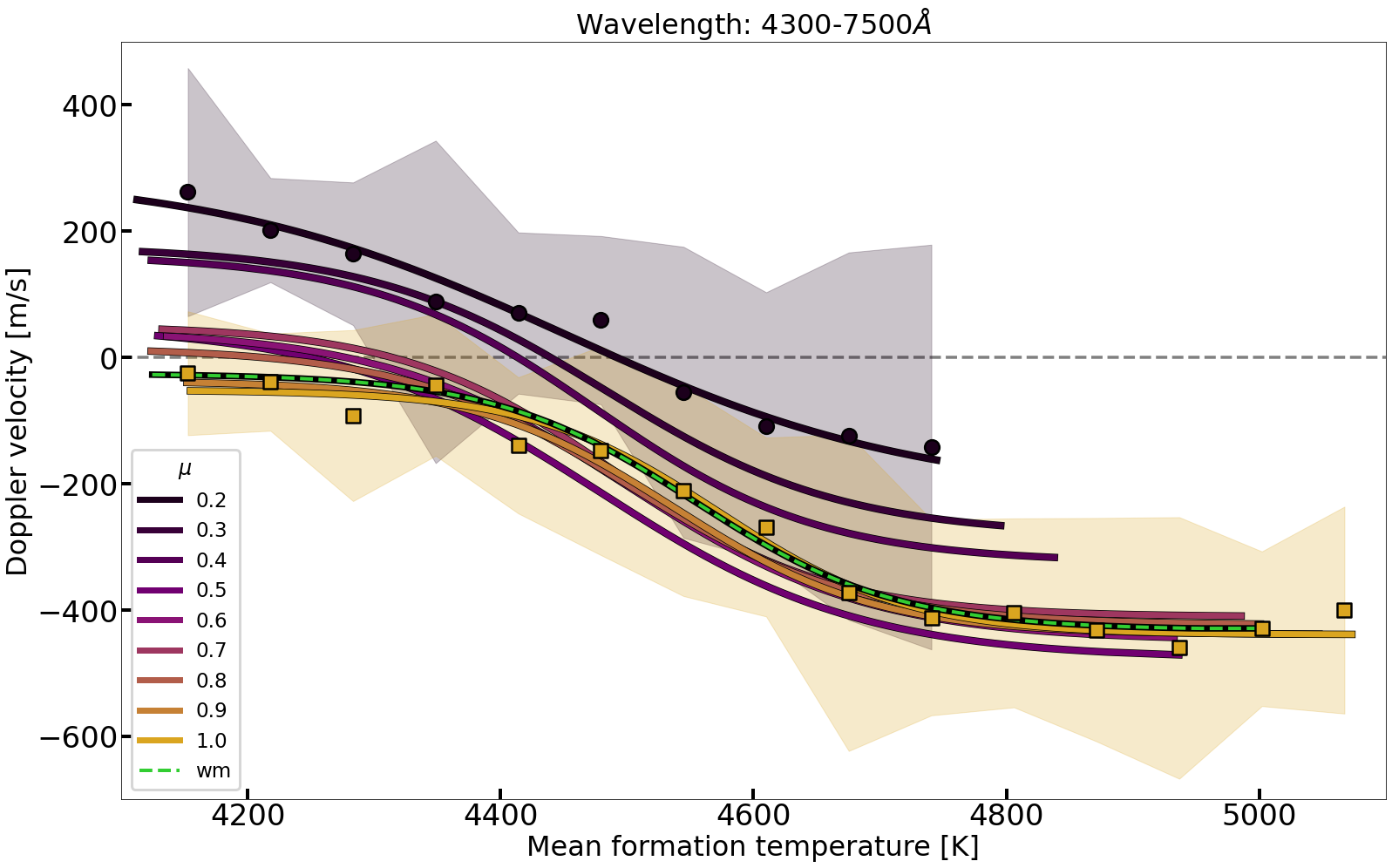}
        \caption{Fitted Doppler velocities, as a function of the mean formation temperature of \ion{Fe}{I} lines, for various $\mu$ positions and for the weighted mean. For $\mu$\,=\,1.0 and $\mu$\,=\,0.2, we show binned data points in 65\,K chunks. The shaded areas display the standard deviation between the fit and the bins.}
        \label{fig:temp_velo}
\end{figure*}

Solar absorption lines are formed over different heights of the atmosphere and therefore over different temperature regions. Since the solar photosphere has a well-known and well-defined temperature gradient, the average formation temperature provides a more accurate indicator of formation heights than the line depth. For an average solar line, the contribution of absorption is defined over a certain temperature region of hundreds of kelvin. 
\cite{2022A&A...664A..34A} used PySME to obtain the average formation temperature of hundreds of lines and plotted those against the convective blueshift of the integrated Sun observed by HARPS-N. The authors used a variety of line species for their investigations and found a trend of decreasing convective blueshift for decreasing temperatures. Decreasing temperatures in the solar photosphere imply higher layers in the atmosphere.
As a next step of our investigation, we examined the Doppler velocities over the solar disc regarding their mean formation temperature. We used PySME (as described in Section \ref{pysme}). With PySME, we compute the cumulative contribution function, which contains the emergent flux at certain optical depths. To estimate the mean formation temperature of a line, we used the temperature where the cumulative contribution function is equal to 50\%. This provides us with the temperature where half of the emission of a line had happened, which serves as average formation temperature. We calculated the respective temperature for each absorption line under consideration of $\mu$.
Figure~\ref{fig:temp_velo} shows the fitted Doppler velocities plotted as a function of the corresponding mean formation temperature. We fit sigmoid functions to our data to reveal trends that would otherwise be hidden among the abundance of data points. The corresponding coefficients are listed in Table~\ref{tab:poly2}. As before, we compare the results of nine $\mu$ positions on the solar disc and calculate the weighted average to all $\mu$ values. Additionally, we plotted binned data points in chunks of 65\,K for $\mu$\,=\,1.0 and 0.2, together with their standard deviations. We can  directly observe the temperature areas observed from the centre to limb. The limb spectra of $\mu$\,=\,0.2, for example, do not contain lines with higher formation temperatures than 4750\,K. 

As general trend, we observe increasing velocities (less blueshift) for decreasing temperatures. This behaviour can be observed for all $\mu$ values. Since higher temperatures occur closer to the bottom of the photosphere where the convective pattern is strongly pronounced, we expect the strongest blueshift in this temperature region. The smaller the temperature, the higher up in the photosphere the lines are formed and, thus, less of the convective overshoot is noticeable. For spectra taken at the disc centre, the average line velocity ranges from about $-435$\,m\,s$^{-1}$ at higher temperatures to approximately $-60$\,m\,s$^{-1}$ for lower temperatures. A similar trend was observed by \cite{2022A&A...664A..34A} for the integrated Sun; although their trend, which involved different species, did not exhibit a sigmoid shape. We notice that the sigmoid shapes get steeper for higher $\mu$'s, which indicates a faster change of Doppler velocities towards the disc centre. This can be explained by the steeper observing angle towards the solar atmosphere. When observing the solar limb, we are looking through the atmosphere at a shallower angle, resulting in a longer observation of certain atmospheric layers compared to observing orthogonal to the surface. \cite{1985SoPh...99...31B} investigated the Doppler velocities of 20 \ion{Fe}{I} lines and used an LTE model atmosphere to calculate the corresponding formation depths. In Fig.~1. of that paper, he shows the Doppler velocities of those lines as a function of formation depth. He used only strong lines which are predominantly in the higher atmosphere and applied the atmospheric model of \cite{1974SoPh...39...19H} to determine log~$\tau$. \cite{1985SoPh...99...31B} fitted an exponential function to represent the trend of his data. Starting from hotter formation regions, the velocity increases from -400 m/s (after subtracting the gravitational redshift) up to about 50\,m\,s$^{-1}$, which represents the behaviour of the stronger lines of our dataset.  When translating the depths into temperatures (using the model of \cite{1974SoPh...39...19H}, we can observe that this temperature scale is stretched compared to ours, with a region from about 4250\,K to 5250\,K.

The functions in Fig.~\ref{fig:temp_velo} do not exhibit the strongest blueshift at the disc centre for a wide range of formation temperatures. In particular, in the temperature range between 4350\,K to 4650\,K, the average convective blueshift observed from $\mu$ positions of 0.5 to 0.9 is stronger than at the disc centre. In fact, we find the strongest blueshift for temperatures higher than 4350\,K for observations at $\mu$\,=\,0.5. A similar trend was observed, for instance, by \cite{2019A&A...624A..57L}, who found that several lines do not exhibit the strongest blueshift at disc centre. Lines from limb positions show velocity shifts with hundreds of m\,s$^{-1}$ more blueshifted. It is possible that this limb effect is due to the surface projection towards the solar limb, since the observations include different movements from various atmospheric heights.
 
For the lowest temperatures in Fig.~\ref{fig:temp_velo}, the average velocities observed at the solar centre show the strongest blueshift. Generally, as shown in Fig.~\ref{fig:dep_veloall}, we observe the smallest blueshifts at the solar limb. For the observations of $\mu$ = 0.2 the average absorption line at formation temperatures lower than 4500\,K is redshifted. The temperature where the redshift occurs drops towards the disc centre but preserves for every $\mu$ smaller than $\mu$\,=\,0.8. The weighted mean of all velocities also ends up entirely blueshifted up to $-35$\,m\,s$^{-1}$ and close to the results of $\mu$\,=\,1.0.


\begin{table}
        \centering
        \small
        \caption{Coefficients of the sigmoid function.}
        \begin{tabular}{@{}rrrrr@{}}
                \hline
                \hline
                $\mu$ & a & b & c & d\\
                \hline
                0.20 & -5.1038e+02 & 7.1109e-03 & 4.4513e+03 & 2.9136e+02\\
                0.30 & -4.5840e+02 & 1.0675e-02 & 4.4834e+03 & 1.7615e+02\\
                0.40 & -4.8749e+02 & 1.1253e-02 & 4.4749e+03 & 1.6251e+02\\
                0.50 & -5.2857e+02 & 9.6980e-03 & 4.4783e+03 & 5.1230e+01\\
                0.60 & -4.9298e+02 & 1.0431e-02 & 4.4994e+03 & 4.4043e+01\\
                0.70 & -4.6185e+02 & 1.1859e-02 & 4.4899e+03 & 5.0932e+01\\
                0.80 & -4.3968e+02 & 1.1247e-02 & 4.5023e+03 & 1.5771e+01\\
                0.90 & -4.0213e+02 & 1.2605e-02 & 4.5372e+03 & -3.6821e+01\\
                1.00 & -3.8750e+02 & 1.3365e-02 & 4.5728e+03 & -5.1730e+01\\
                \\
                wm & -4.0633e+02 & 1.2565e-02 & 4.5537e+03 & -2.5677e+01 \\
                \hline
        \end{tabular}
    \tablefoot{Coefficients of the sigmoid function $ a / (1 + \exp(-b *(x - c))) +d$ to describe the trends of the Doppler velocity as a function of formation temperature as plotted in Fig.\,\ref{fig:temp_velo}}
\label{tab:poly2}
\end{table}  

\begin{figure}
        \centering
        \includegraphics[width=0.49\textwidth]{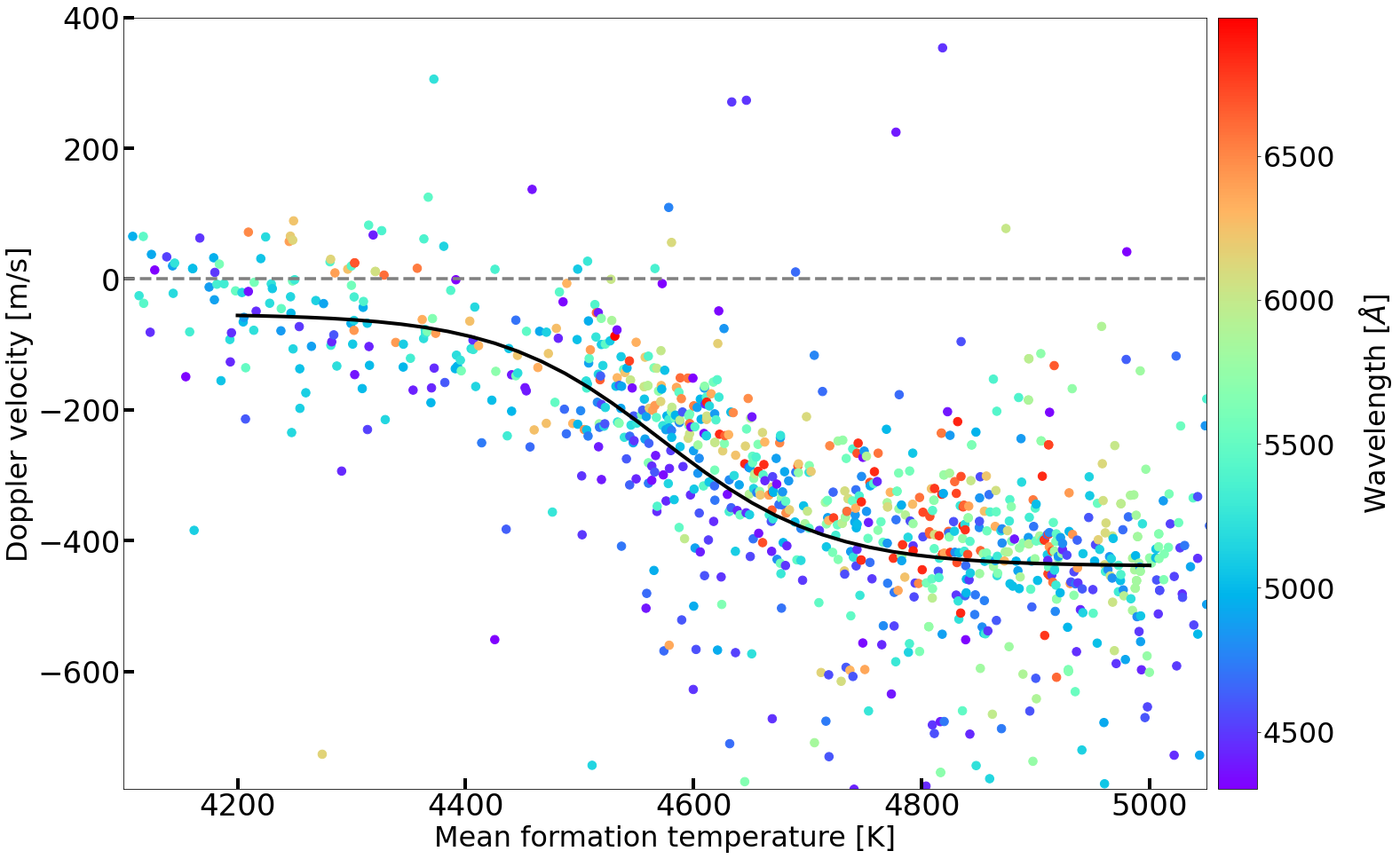}
        \caption{Doppler velocity of \ion{Fe}{I} line cores for $\mu$ = 1.0 as a function of mean formation temperature, colour-coded over wavelengths. The fit corresponds to the fit from Table~\ref{tab:poly2}.}
        \label{fig:velo_temp_wl}
\end{figure}

Due to the wavelength impact of the Doppler velocities as a function of line depth we also chose to investigate the velocities as a function of formation temperature. Figure~\ref{fig:velo_temp_wl} shows the Doppler velocities over the mean formation temperature colour-coded over wavelengths for $\mu$=1.0. The lines follow a well-defined relation with a remaining trend over wavelengths. However, this wavelength dependence is less pronounced compared to the results in Fig.~\ref{fig:velo_depth_wl}. We observe that the average velocities of absorption lines at shorter wavelengths show still more blueshift than lines at longer wavelengths. This velocity trend could be a remnant of the line formation itself. Since shorter wavelengths are formed lower in the atmosphere because of the lower continuum's opacity, they show a greater blueshift in general as they origin with greater stellar surface contrast.

All in all, our mean formation temperatures result from simplifications of an exceedingly more complex reality.
The velocity contribution function of lines over the photospheric height differs a lot in width and shape for different lines \citep{1979SoPh...61..251D, 2002A&A...381..253B}.
Nevertheless, the way we determined the mean formation temperature provides a convincing initial estimate. Our results give an insight into the contribution of different temperature regions in the atmosphere to the convective blueshift, which can be used for further investigations and simulations of the solar atmosphere.

\subsection{Convective blueshift over line depth utilising mean formation temperature}

In Section~\ref{lintemp}, we determined the dependence of Doppler velocities on mean formation temperatures and showed that this relation is applicable to lines at all wavelengths. This can be used to attempt reverse engineering of the relation between Doppler velocity and line depth, to which the dependence on wavelength adds significant scatter as shown in Fig.~\ref{fig:velo_depth_wl}. Our assumption is that the relations seen in Fig.~\ref{fig:temp_velo} reflect the pattern of convective motion (independent of wavelength), and that we can use these relations to reconstruct the dependence between Doppler velocity and line depth as a function of limb position and wavelength.

\begin{figure}[ht]
    \centering
    \includegraphics[width=0.49\textwidth]{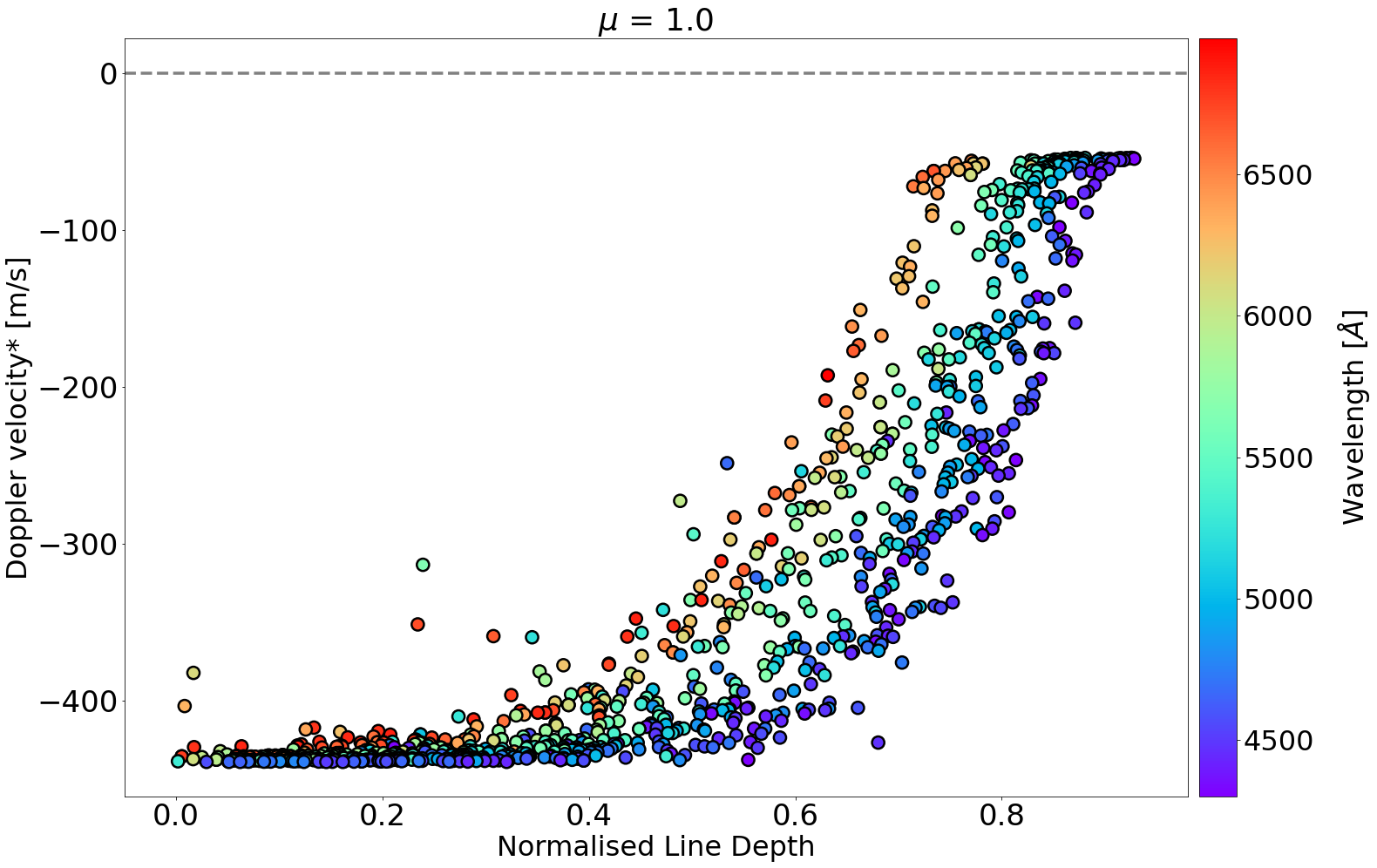}
    \includegraphics[width=0.49\textwidth]{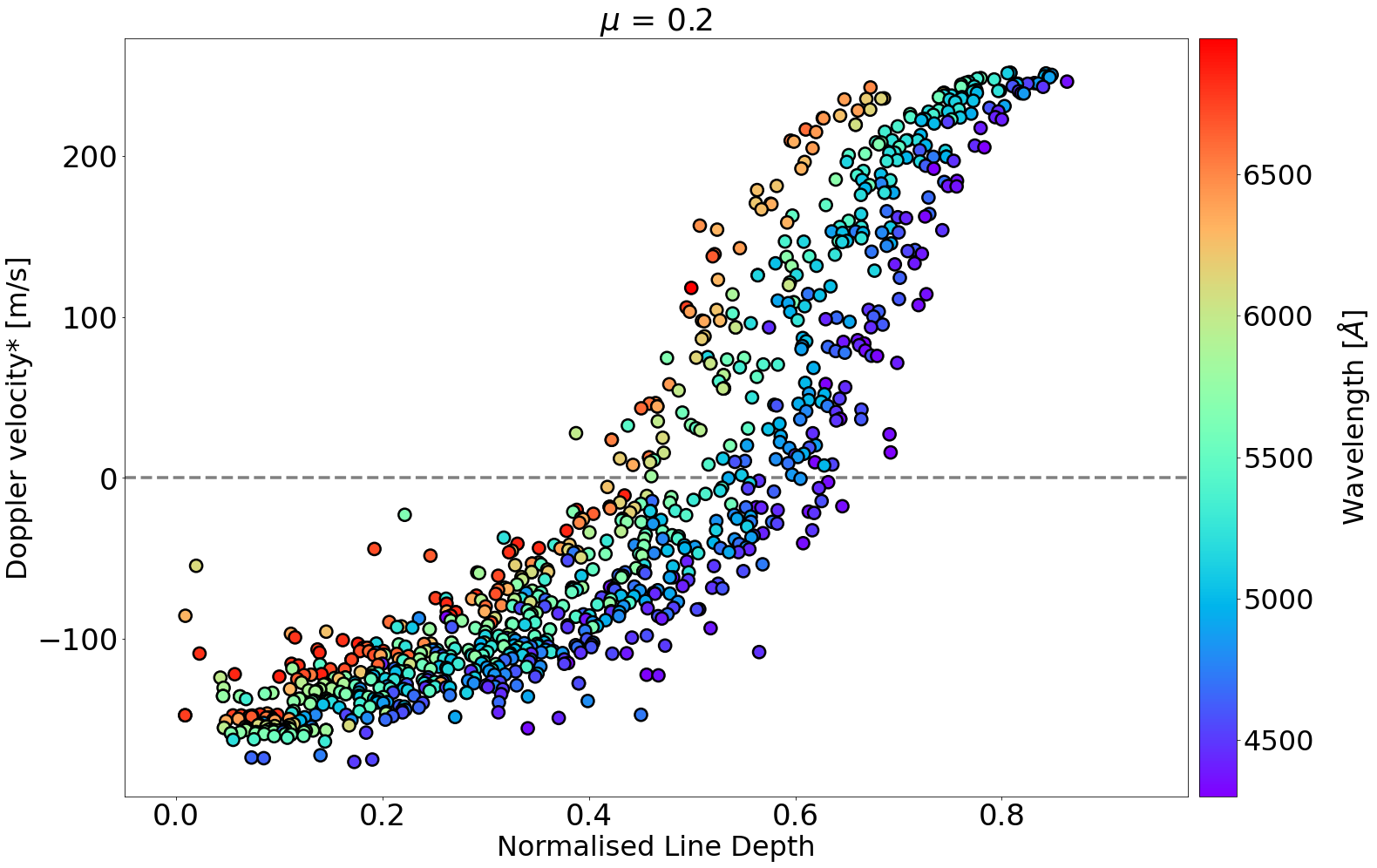}
    \caption{Doppler velocity corresponding to the sigmoid-shaped fit of Fig.~\ref{fig:temp_velo} as a function of the line depth of \ion{Fe}{I} lines for $\mu$ = 1.0 and $\mu$ = 0.2.}
    \label{fig:dep_velo_reversed}
\end{figure}

We use the relations as shown in Fig.~\ref{fig:temp_velo} and Table~\ref{tab:poly2} to determine the Doppler shift for each individual line and limb position. Figure~\ref{fig:dep_velo_reversed} shows the results of this approach for disc centre and limb position of $\mu$\,=\,0.2. A comparison between Fig.~\ref{fig:dep_velo_reversed} and the velocity distribution of Fig.~\ref{fig:dep_veloall} shows that the reconstructed distribution for $\mu$\,=\,1.0 (Fig.~\ref{fig:dep_velo_reversed}) does not exhibit a velocity minimum at line depths of 0.2\,-\,0.4. We conclude from this that the minima seen in Fig.~\ref{fig:dep_veloall} and Fig.~\ref{fig:dep_velo_b}  (and in earlier works) were introduced by the choice of a polynomial and distribution of lines and line depths, but this is not a feature of the convective pattern.

The dispersion of Doppler velocities for $\mu$\,=\,1.0 in Fig.~\ref{fig:dep_velo_reversed} is particularly small, that is, we observe a rather constant distribution of velocities in shallow lines at disc centre. A possible reason for the stronger dispersion of shallow lines in Fig.~\ref{fig:velo_depth_wl} is the more challenging measurement of the line position due to continuum noise. In Fig.~\ref{fig:dep_velo_reversed}, the dispersion of velocities only starts at a line depth of 0.3 and deeper. As we can see, this dispersion is due to the wavelength dependence of the line depths and is also a dominant factor in the plot for $\mu$\,=\,0.2. This wavelength dependence arises from the varying effective absorption and changing path lengths through the photosphere, as discussed in Chapter \ref{lindepth}. The velocities can vary by several hundred meters per second depending on the specific wavelength range within the spectrum, which is of the same order as the dispersion given in Fig.~\ref{fig:velo_depth_wl}. The strong wavelength dependent variation of Doppler velocities highlights the advantage of studying velocities in terms of formation depths or formation temperatures, rather than line depths.

\section{Summary} \label{Summary}

In this paper, we investigated the centre-to-limb variation of more than 1000 solar \ion{Fe}{I} lines with respect to the convective motions of the quiet Sun. We used our resolved solar atlas \citep{2023A&A...673A..19E} to examine the behaviour of Doppler velocities for nine observing positions, $\mu,$ of the solar disc. These centre-to-limb positions allow us to examine changing viewing angles towards the solar surface.
To investigate the convective blueshift of the solar atmosphere, we first examined the Doppler velocities of line positions with respect to their line depths. Below, we list our main findings:
\begin{enumerate}
    \item Overall, the absorption lines of the spectra towards the solar limb show decreasing convective blueshift. This decrease is a direct effect of observing higher atmospheric layers, with less granular contrast, towards the solar limb. The averaged velocities of all limb observations show a redshift for the strongest lines. This redshift arises at increasingly shallower lines as the observations were taken closer towards the solar limb. This shift from blueshift towards a redshift of lines observed from centre-to-limb has been noticed on single lines before \citep{2018A&A...611A...4L, 2019A&A...624A..57L}. At $\mu$\,=\,0.9 and the disc centre ($\mu$\,=\,1.0), the averaged velocities remain blueshifted for every line depth.
    
    \item We calculated the weighted average of all Doppler velocities and compared it with the Doppler velocities we obtained by examining the integrated IAG Sun spectrum. Comparing both, we notice that the results are tilted to each other. The polynomial of the weighted mean shows a shallower increase of velocities towards deeper lines than the fit for the integrated Sun. Both fits cross each other at the line depth of 0.7. The integrated results show less convective blueshift of 70\,m\,s$^{-1}$ for strong lines and stronger convective blueshift of 40\,m\,s$^{-1}$ for shallower lines.

    \item We inspected the wavelength dependency of line depth and the corresponding convective blueshift. The wavelength sensitivity of the Doppler velocities increases with line depth at every $\mu$ position.
\end{enumerate}
Since formation temperatures serve as a better proxy for geometrical formation height than line depth, we additionally calculated the mean formation temperature of the \ion{Fe}{I} lines. To obtain the temperatures, we computed the contribution function of spectral synthesis by PySME. We reached the following conclusions:

\begin{enumerate}
\setcounter{enumi}{3} 
    \item The general behaviour of the Doppler velocities over formation temperatures exhibits similar behaviour as for line depths. We obtained decreasing blueshift for decreasing formation temperature for every $\mu$, as a higher temperature is equivalent to deeper in the photosphere where convection is strongest. The averaged Doppler velocities can be described by a sigmoid function, which gradually becomes steeper for larger $\mu$'s.
    \item The observed convective blueshift is smallest close to the solar limb. Velocities even exhibit a redshift at lower temperatures for all observed $\mu$ positions lower 0.8. The redshift of the average line occurs earlier the further at the limb the observation derived. 
    \item The convective blueshift is not strongest at the disc centre over a large range of formation temperature. At $\mu$ positions of 0.5 till 0.9 we receive higher convective blueshift velocities than for the disc centre over the temperature range of about 4350\,K to 4800\,K. At $\mu$\,=\,0.5, we observe the strongest blueshift.
    \item We found a remaining wavelength dependency of the Doppler velocities over formation temperature. One reason for this remnant of wavelength dependency could be the varying continuum's opacity, which causes shorter wavelengths to form deeper in the solar atmosphere where the surface contrast is larger.
    \item We used the sigmoid function we obtained from the velocity fit as a function of formation temperature and determined the corresponding line depth to reverse engineer Fig.~\ref{fig:dep_veloall}. We observed a scatter that only originates from the wavelength dependency.
\end{enumerate}
In conclusion, the convective profiles and comprehensive convective information presented in this study hold the potential to enhance solar and stellar atmospheric models, leading to a deeper understanding of these complex systems. It would be educational to probe simulations based on their convective behaviour and compare these results to the outcomes of the present work.

\begin{acknowledgements}
        We thank the referee for very constructive suggestions and ideas. This work has made use of the VALD database, operated at Uppsala University, the Institute of Astronomy RAS in Moscow, and the University of Vienna. ME acknowledges financial support through the SPP1992 under Project DFG RE 1664/17-1.
\end{acknowledgements}

\bibliographystyle{aa} 
\bibliography{Ellwarth} 
\clearpage

\begin{appendix}

\section{Deviations between line lists} \label{deviation}

To determine the absolute convective blueshift of absorption lines, we need to establish the reference wavelengths. We counter-checked the \ion{Fe}{I} line positions of different lists against each other. The wavelength range considered is $4200-8000$\,\r{A}. Many studies have already provided theoretical line positions and since \ion{Fe}{I} lines have been extensively studied in the astrophysical context, their properties, such as their rest wavelengths, have been extensively examined. The line lists we consider here are those of \cite{1994ApJS...94..221N} and \cite{K14} obtained from the VALD3 catalogue, along with the re-calibrated \cite{1994ApJS...94..221N} lines received from the NIST Atomic Spectra Database \citep{NIST_ASD}.

\begin{figure}
        \centering
        \includegraphics[width=0.49\textwidth]{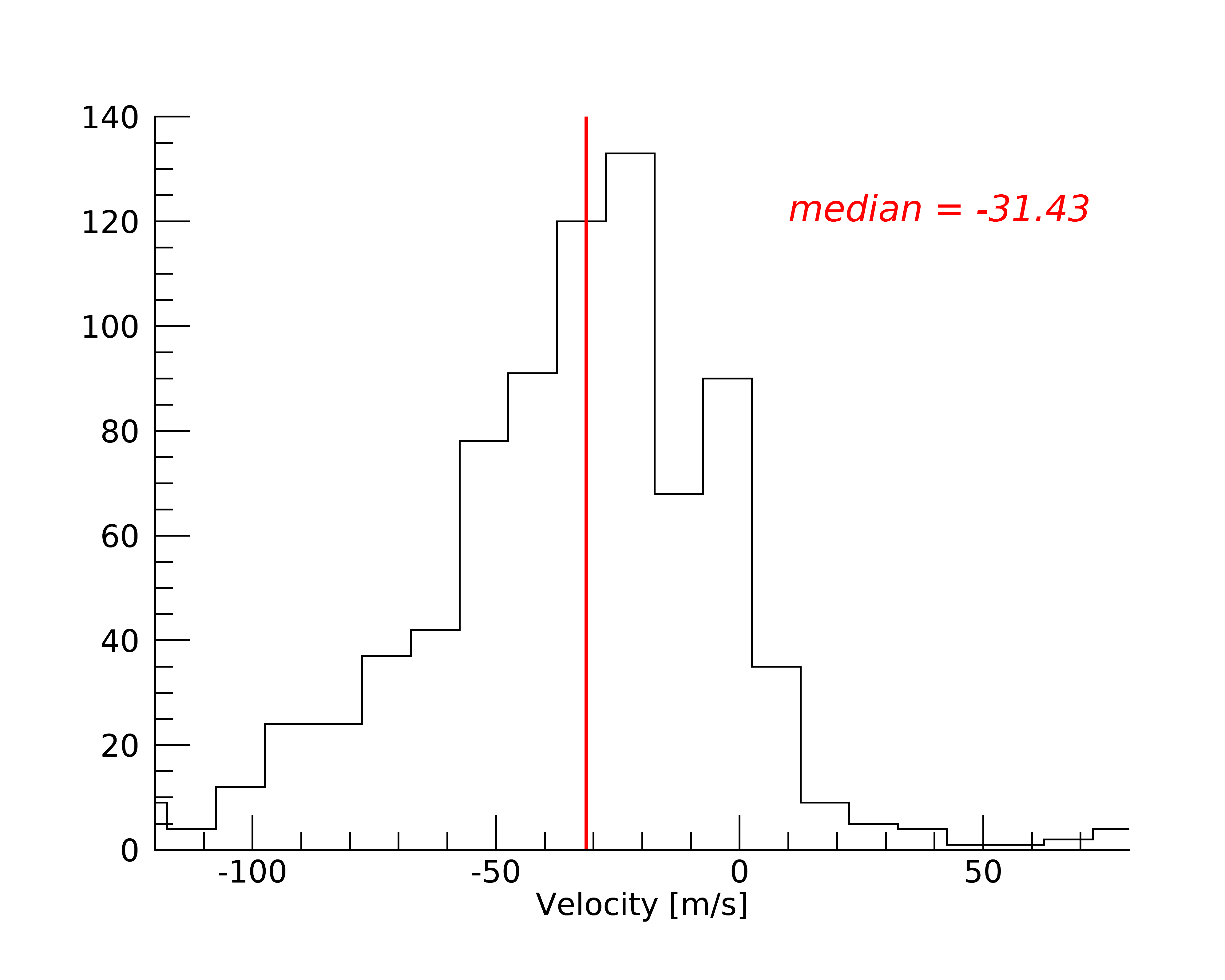}
        \caption{Histogram of velocities regarding to the deviation of the \ion{Fe}{I} line lists of N94 and VALD3.}
        \label{fig:hist_velo}
\end{figure}

The line positions ($\lambda_N$) from the \cite{1994ApJS...94..221N} line list, hereafter referred to as N94, were first compared against the line positions from the VALD3 line list ($\lambda_V$). The difference between the two lists in velocity units is expressed by
\begin{align}
    \Delta v = c \cdot \frac{\lambda_V - \lambda_N}{\lambda_N},
\end{align}
where c is the speed of light. The histogram in Fig.~\ref{fig:hist_velo} displays the distribution of the velocity difference of 980 \ion{Fe}{I} lines. For this plot we used lines with a line depth greater or equal than 0.4. The median velocity difference of $\Delta v$ = -31.43\,m\,s$^{-1}$ indicates that the line positions in VALD3 are generally shifted towards shorter wavelengths compared to the positions in the N94 list. Changing the line depth threshold has barely an impact on the median velocity, with a range of only $\pm 1$\,m\,s$^{-1}$. The scatter of the shifts are mainly spread over a range of about 100\,m\,s$^{-1}$.

\cite{2021A&A...654A.168L} previously reported a discrepancy between VALD3 and N94 when investigating the convective blueshift of \ion{Fe}{I} lines, but they attributed this mismatch predominantly to the deep lines. In our analysis, we have identified a systematic mismatch between the two line lists that seems independent of line depth.
\cite{2020A&A...643A.146G} compared the N94 line positions with the \ion{Fe}{I} wavelengths of the NIST Atomic Spectra Database. NIST contains re-calibrated N94 wavelengths. The authors found that the NIST lines are at shorter wavelengths by about 20 and 35\,m\,s$^{-1}$ (observed and ritz wavelength) compared to the un-calibrated N94 lines. We also checked the NIST lines against our N94 list and received a mean difference of shorter wavlenghts for NIST of 16\,m\,s$^{-1}$ for the observed wavelengths and 29\,m\,s$^{-1}$ for the ritz wavelengths. This smaller deviation of N94 and NIST may arise because we pre-filter our lines using the unblended \ion{Fe}{I} list of \cite{1989A&AS...77..137T, 1990A&AS...82..179T}. Therewith VALD3 and NIST agree by 2.4\,m\,s$^{-1}$.


\end{appendix}

\end{document}